\documentclass[twocolumn,letterpaper]{IEEEAerospaceCLS}  

\usepackage[]{graphicx}    
\newcommand{\ignore}[1]{}  
\usepackage[noadjust]{cite}
\usepackage{amsmath}
\usepackage[ruled,lined,linesnumbered]{algorithm2e}
\usepackage{siunitx}
\usepackage[hyphens]{url}
\usepackage{hyperref}  
\usepackage{tabularx}
\usepackage{flushend}

\usepackage[capitalize]{cleveref}
\begin{document}

\title{Uncovering GNSS Interference with Aerial Mapping UAV}

\author{%
Marco Spanghero\\
\textit{Networked Systems Security Group} \\
\textit{KTH Royal Institute of Technology}\\
Stockholm, Sweden \\
marcosp@kth.se
\and 
Filip Geib\\
\textit{Wingtra AG}\\
\textit{Zürich, Switzerland}\\
filip.geib@wingtra.com
\and 
Ronny Panier\\
\textit{Wingtra AG}\\
\textit{Zürich, Switzerland}\\
ronny.panier@wingtra.com
\and 
Panos Papadimitratos \\
\textit{Networked Systems Security Group} \\
\textit{KTH Royal Institute of Technology}\\
Stockholm, Sweden \\
papadim@kth.se
}

\maketitle

\thispagestyle{plain}
\pagestyle{plain}

\begin{abstract}
Global Navigation Satellite System (GNSS) receivers provide ubiquitous and precise position, navigation, and time (PNT) to a wide gamut of civilian and tactical infrastructures and devices. Due to the low GNSS received signal power, even low-power radiofrequency interference (RFI) sources are a serious threat to the GNSS integrity and availability. Nonetheless, RFI source localization is paramount yet hard, especially over large areas. Methods based on multi-rotor unmanned aerial vehicles (UAV) exist but are often limited by hovering time, and require specific antenna and detectors. In comparison, fixed-wing planes allow longer missions but are more complex to operate and deploy.
A vertical take-off and landing (VTOL) UAV combines the positive aspects of both platforms: high maneuverability, and long mission time and, jointly with highly integrated control systems, simple operation and deployment.
Building upon the flexibility allowed by such a platform, we propose a method that combines advanced flight dynamics with high-performance consumer receivers to detect interference over large areas, with minimal interaction with the operator. The proposed system can detect multiple interference sources and map their area of influence, gaining situational awareness of poor GNSS quality or denied environments. Furthermore, it can estimate the relative heading and position of the interference source within tens of meters. The proposed method is validated with real-life measurements, successfully mapping two interference-affected areas and exposing radio equipment causing involuntary in-band interference.
\end{abstract} 

\maketitle

\thispagestyle{plain}
\pagestyle{plain}

\tableofcontents

\section{Introduction}

Global Navigation Satellite Systems (GNSS) receivers are integrated in a wide gamut of systems. Nevertheless, due to the open signal specification allowing potential adversaries to craft valid GNSS signals and the current lack of cryptographic features, civilian GNSS receivers are susceptible to a wide range of disturbances. Attackers can transmit signals spanning from involuntary spurious emissions causing in-band interference to advanced malicious manipulation targeting the PNT solution \cite{psiaki2016gnss, Lenhart2022, HumphreysAssessingSpoofer}. Among all attacks, jamming is the simplest attack to mount, denying GNSS reception in a large area with relatively simple equipment, and it is an effective stepping stone for other attacks. 

Given the low complexity required for a successful GNSS jamming attack, recent events show that portable jammers are not only available but also extremely effective \cite{CNETNewark, insideGNSSNewark, SkytruthJamming}. Extensive risk analysis \cite{john2001vulnerability,Mitch2011} show that small handheld devices can produce powerful and sustained interference blocking the GNSS signal reception, and effectively impairing the receiver. Because of the low SWaP (Size, weight, and Power) portable jammers are easily concealed, making localization of the interference source challenging, especially for multiple transmitters in a large area.  Although mitigation is possible \cite{Borio2012}, the attacker can remain effective even if receiver-based countermeasures are in place, with the only solution being localization and physical removal of the interference source (either for active disturbance devices or the byproduct of non-compliant radio transmitters) Mapping of the affected area is necessary to understand the extend of the disturbance and localization of the transmitting device. Traditional interference hunting techniques, relying on survey antennas, and manual measurements done by ground operators are effective in localizing interference sources. However, operations are often restricted to small areas, rely on the ability and experience of the operator, and are generally suited for short-range measurements \cite{Anritsu_appnote_rfi,Keysight_appnote_rfi}. Mapping interference sources over a large area is complex, costly, and time-consuming. Compared to a ground vehicle or operator, a UAV allows one to explore larger areas, and while the benefits of multi-rotor platforms are clear, there are limitations to the flight time and efficiency achievable. The recent development of consumer unmanned aerial vehicles with vertical take-off and landing capability (UAV-VTOL) paved the road to deploying improved flight dynamics, combining the high efficiency of a plane with the simplicity of deployment of a multi-rotor drone.

In this work, we explore exactly this: how to combine the advanced maneuverability of a VTOL UAV with a consumer GNSS receiver to perform radio frequency interference (RFI) scans over large areas, allowing the UAV to establish GNSS quality of service (QoS) zones, automatically. We show how few scans at selected vantage points allow the UAV to obtain full situational awareness over the interference-affected area, pinpointing the location of the jammer and its area of influence, with minimal operator intervention. Measurements performed in a real-life scenario show that the detection of multiple sources of interference is possible within \SI{25}{\meter} from the true position of the interference transmitter(s).

The rest of the paper is organized as follows: \cref{section:related_work} discusses the current state of the art and the existing applications of VTOLs in RFI mapping, \cref{section:attacker-model} presents the type of interference transmitter considered in the scope of this work, \cref{section:methodology} gives an overview of the method used to measure and map interference, \cref{section:experimental_setup} described the setup used to validate the method, \cref{section:results} shows the achieved results and performance and \cref{section:conclusions} discusses possible future work and final conclusions.

\section{Related Work}
\label{section:related_work}

Detection and classification of jammers and interference sources using small, portable low-cost devices is possible \cite{Isoz2010} but localization is often beyond the scope of single antenna detection devices, lacking direction-finding capabilities unless networks of cooperative agents provide measurements for that purpose (\cite{olsson2022participatory, olsson2023participatory,Lindstrom2007,Strizic2018,Cetin2014}). On the other hand, participatory schemes are subject to different vulnerabilities that are beyond the scope discussed in this work and need to rely on dedicated security and privacy preserving systems \cite{GisdakisGP:J:2016, GisdakisGP:C:2015}.

Methods using fixed stations to monitor interference and geo-locate its source(s) exist \cite{Poncelet2012,Akos2012,Spens2022,Hashemi2019}, but are limited to a fixed area and cannot be deployed upon need, or easily moved to a new area. Techniques that rely on direction finding exist, leveraging multi-element antennas \cite{Montgomery2009,Magiera2015,Bhatti2012,Bours2014} and more specifically Controlled Path Reception Antenna (CRPA) receivers allow precise localization and elimination of the interference source \cite{Caizzone2019,Perez-Marcos2023,Marcos2018}. Nevertheless, such methods require specialized hardware and complex antenna systems and are not suitable for consumer devices, or are unfeasible to install on small aerial vehicles (due to size, power consumption, and computational requirements). 

More recent examples relying on space-based jammer detection and geo-localization (\cite{Clements2022,Clements2023}) show very promising results but require dedicated infrastructure and capabilities that are even further beyond the scope of consumer-grade interference surveying. Few examples of airborne jammer detection and localization systems exist and rely on multi-rotor vehicles \cite{Spicer2015,Perkins2016}, limiting the range achievable at mission time. Improvements based on multi-element antenna systems show that such a method is capable of precisely localizing interference transmitters \cite{Perkins2019}, but requires the use of specialized receivers, antenna systems, and UAV platforms. In comparison, the use of a VTOL plane instead of a multi-rotor drone allows for longer reach due to the efficient flight dynamics, at the cost of slightly diminished localization accuracy, which proves to be a reasonable trade-off. A major drawback is that VTOL planes are generally bigger and more cumbersome to transport compared to multi-rotor platforms. Nevertheless, multi-rotor configurations that allow comparable flight time are similar in size if not bigger than VTOL planes \cite{matrice600}. Moreover, fixed-wing planes are often more complex to control for inexperienced operators compared to multi-rotor ones; the complexity can be lowered by employing sophisticated fly-by-wire controllers. Additionally, the approach considered in this work aims at re-using the components within an existing UAV system and exploiting advanced flight dynamics to perform localization of the interference source. This is possible by using the same GNSS receiver for navigation and RFI detection. The approach presented in \cref{section:methodology} can be leveraged by any UAV with suitable flight dynamics, independently of the receiver platform.

\section{Interference/Jamming model}
\label{section:attacker-model}

Not all interference emissions are necessarily due to malicious intent: several cases exist of spurious emissions from low-quality or non-compliant devices. Although these devices cannot be deemed adversarial, they still represent a sensible threat to GNSS receivers and for this reason, are worth considering within the scope of the interference model. As an example, GPS L1 is susceptible to second-order harmonics of transmitters operating in the \SI{781}{\mega\hertz} - \SI{794}{\mega\hertz} bands (TV broadcast in North America) and third-order harmonics in the \SI{521}{\mega\hertz} - \SI{530}{\mega\hertz} bands. Furthermore, according to the most recent Frequency Orientation plan released by the Swedish Post and Telecom Authority the range \SI{1240}{\mega\hertz} - \SI{1300}{\mega\hertz}, corresponding to G2 (Glonass L2) and E6 (Galileo high accuracy services) is shared with radio amateur operators who can operate license-free in this channel with equivalent isotropic radiated power up to \SI{200}{\watt} \cite{PTSSpectrum}. While these allocations tend to be country- and regulation-specific, meaning not every area is affected to the same extent, it is clear how simple coexistence of radio standards can affect GNSS robustness.

On the other hand, the voluntary transmission of interference signals is an effective and cost-effective attack vector that can be used as a stand-alone denial tool or to bootstrap more complex scenarios. We consider a static (or slowly moving) interference source that aims at producing enough disturbance in the radio channel to lower the legitimate carrier-to-noise ratio (C/$N_0$). The attacker is capable of producing an arbitrary type of interference: modulated (mimicking legitimate satellite PRN) or unmodulated (flat band noise or chirp) interference with varying duty cycles and power. The former is common for small portable devices with power limited to a few 100mW up to 1W, which are readily available to the general public and for this reason, are the most likely to be found in the field. Additionally, the attacker could employ one or multiple transmitters at different locations possibly cooperating to maximize the effectiveness of the jamming signal. Normally, such an attacker does not adopt any specific strategy to impair the GNSS receiver but pattern-based jammers (interrupting the signal at a specific time, \cite{Curran2016}) have been successful in disrupting GNSS signal quality. Nevertheless, this requires additional logic often absent in low-tier transmitter devices.

\section{Methodology}
\label{section:methodology}
\begin{figure}[h]
\centering
\includegraphics[width=\columnwidth]{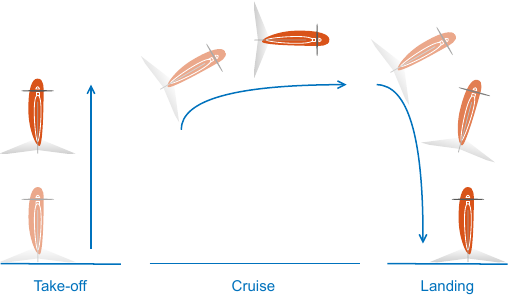}
\caption{Transition between different flight modes.}
\label{figure:drone-transition}
\end{figure}

For long-range wide-area scans as the ones presented in this work, the use of a VTOL platform is preferable. Two operational modes, combining the advantages of each aforementioned UAV platform, are defined: cruise flight mode and hover mode, shown in \cref{figure:drone-transition}. In cruise mode, the UAV flies in the assigned space, completing the configured mission: this extends the range of the operations compared to a multi-rotor system, allowing the UAV to cover larger areas during a single mission. In hover mode, the UAV maintains its position, flying vertically, like a multi-rotor platform: this allows the operator to perform on-the-spot operations like pointing the survey antenna to a specific heading.

Intuitively the method is as follows: the UAV surveys the RF spectrum at different pre-determined locations, scanning the interference level at each antenna heading, here called "horizon scanning". The collected measurements are fused in a global view that combines the information of the individual scans in an overlay map, highlighting simultaneously the extent of the interference-affected area and the likely position of the interference transmitter, without the UAV entering the interference-affected area.

\begin{figure}
	\centering
	\includegraphics[width=\columnwidth]{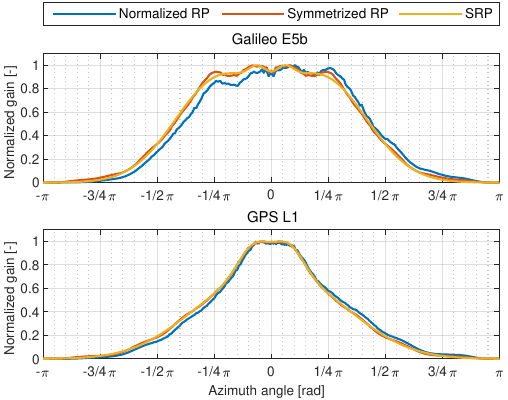}
	\caption{Receiver antenna radiation pattern at different frequencies used to ensure consistent performance.}
	\label{figure:gain-mask}
\end{figure}

\textbf{Antenna compensation} - Scanning for interference uses the same advanced GNSS receiver normally used for cruise mode navigation. During scanning operations, the GNSS receiving antenna's major axis is parallel to the ground, allowing for geometry-based selectivity in addition to the antenna reception pattern. To isolate the specific heading the interference is transmitted from, antenna directivity is important. Notably, the antenna directivity is not guaranteed to be the same over the various frequency bands of operation; it might vary slightly, leading to varying localization performances depending on the band. To ensure a flat gain profile and the same directivity over the entire frequency range we apply a sensitivity mask at the center frequency of each monitored channel. This ensures the same directional sensitivity in each band, at the cost of slight gain compression for those bands whose antenna sensitivity is higher. \cref{figure:gain-mask} shows the calculated gain masks. Based on the measured antenna directivity, the antenna radiation pattern (RP) is first normalized. Then, a set of constraints is enforced (the radiation pattern is symmetric, is maximum along the antenna major axis, and decays to zero towards the backplane) to extract a model of the antenna radiation patterns in a specific band, defined as a symmetrized radiation pattern. Finally, the standard radiation pattern (SRP) is obtained by fitting a Gaussian mixture model to the symmetrized RP.

\textbf{Mission planning} - A flight mission includes several short transitions from cruise to hover mode, providing scans from different positions. The flight path is designed to suit a specific mission and can potentially be updated to take into account the results of the interference detection process. In hover mode, the UAV cannot rely on its main GNSS receiver for navigation, which is used to perform interference monitoring. A secondary GNSS receiver and inertial navigation provide both localization and attitude. Notably, this requires the UAV to operate outside the denied area: long-range sensing is possible due to the high gain of the main scanning antenna, allowing the UAV to detect potential denial of service early on. 

\textbf{Relative power spectral density estimation} - The system collects samples of the raw baseband spectrum and estimates the power spectral density in each channel. The power spectral density (PSD) estimation is obtained as \cref{formula:psd}, where $k$ is the $N$-point Discrete Fourier Transform (DFT) bin number. The measured power density is recorded and the measurement is performed for each sector based on the UAV's heading angle. 

\begin{equation}
 \label{formula:psd}
\begin{split}
	P(f_k) & = \frac{1}{N}\left|\sum_{n=0}^{N-1} x[n]e^{-j2\pi f_k n]}\right|^2 \\
               & = \frac{1}{N}\left|\sum_{n=0}^{N-1} x[n]e^{-j2\pi \frac{k}{N} n]}\right|^2
\end{split}
\end{equation}

Traditionally, systems relying on the power spectral density to localize the source of interference require assumptions on the power of the transmitter. Often such assumptions are hard to satisfy, specifically when dealing with small personal devices due to their variability and inconsistency. Additionally, reliable power measurements require a power-calibrated front-end \cite{Miguel2023} which is beyond the scope of what is available in any GNSS receiver. To solve this issue, our algorithm does not rely on the absolute power estimation of the adversarial transmitter but leverages the UAV's dynamics to perform several scans in different positions and exploit geometrical diversity in the measurements. Observations from different locations create a synthetic view of the RFI environment, relying on the relative power variation more than the absolute received power.  

\textbf{Horizon scan} - At each location, the UAV performs a comprehensive horizon scan using the main GNSS receiver. The heading is incremented in steps equivalent to the antenna directivity and the power density is measured at each heading. An expectation density algorithm (shown in \cref{algo:heatmapfusion}) is used to extract the power density expectation at each antenna heading and considers the following inputs: radiation heatmap $HM$ as heading-referenced power spectrum estimation, location and heading of the UAV $P$ at the scan position, and radiation pattern $SRP$. For each coordinate $[x,y]$ within the heatmap $HM$ (Line 2) the spectral density is decomposed in the angular components (Line 3), and corrected according to the SRP at the survey band (Line 4). This information is extracted at each survey site.

\begin{algorithm}
    \KwData{$HM$, $P=(x^\prime, y^\prime, \psi^\prime)$, $SRP$}
    \KwResult{$HM$}
\Indp\Indp\Indp\Indp\Indp\Indp\Indp\Indp
\Indp\Indp\Indp\Indp
    
    \BlankLine

\SetInd{2.em}{1em}
    $HM \gets 0$\\
    \For{$[x,y] \gets D\{HM\}$}{
        $\Delta\psi \gets \psi^\prime - \arctan(x-x^\prime, y-y^\prime)$\\
        $HM(x,y) \gets SRP(\Delta\psi)$\\
    }
    \caption{Fusion of expectation density.}
    \label{algo:heatmapfusion}
\end{algorithm}

 \textbf{Combination of scans} - The intersection of the expectation densities that shows the highest anomalous power distribution is the area where the interference transmitter is most likely to be located. Practically the process consists of directly projecting the heading-referenced power measurements in the 2-D plane and by accumulation of the power density in each radiation pattern, forming a complete map. The areas with the highest accumulated power are the ones where the transmitter is located. Additionally, the accumulation values can be used to trace the influence area of the transmitter.
The final map $HM_{fuse}$ is fused by stacking together all heatmaps produced by the expectation density algorithm (shown in \cref{algo:heatmapfusion}) for all executed scans. This approach is presented in \cref{algo:mapfusion}, which additionally to \cref{algo:heatmapfusion} considers the set of various locations and headings of the UAV ($P = [P_1,...,P_N]$) at all available $N$ scan positions, defined by the mission. The intermediary heatmaps are accumulated (Line 5) and the final map is then created by normalizing the integrated heatmap by the number of available scans (Line 8). Additionally, a threshold function is applied (not shown in \cref{algo:mapfusion}), to suppress the noise caused by the integration, highlighting the areas where the jammer has the higher likelihood of being located and its coverage zone.

\begin{algorithm}
    \KwData {$HM$ , $P = [P_1,...,P_N]$, SRP}
    \KwResult {$HM_{fuse}$}
\Indp\Indp\Indp\Indp\Indp\Indp\Indp\Indp
\Indp\Indp\Indp\Indp

  \BlankLine

\SetInd{2.em}{1em}
    $HM \gets 0$\\
    \For{$i=0$ \KwTo $N$ }{
        \For{$[x,y] \gets D\{HM\}$}{
            $\Delta\psi \gets \psi^\prime_i - \arctan(x-x^\prime_1, y-y^\prime_1)$ \\
            $HM(x,y) \gets HM(x,y) + SRP(\Delta\psi)$\\
        }
    }
    $HM_{fuse} \gets HM/N$\\
    \caption{Fusion of multiple expectation densities.}
    \label{algo:mapfusion}
\end{algorithm}

\begin{figure}[h]
	\centering
	\includegraphics[width=\columnwidth]{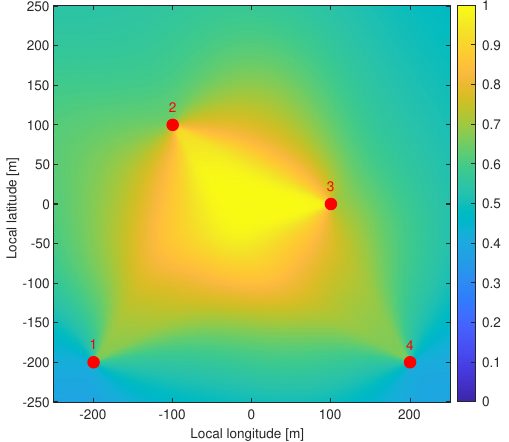}
	\caption{Simulation of one jammer and localization from geometrically advantageous positions.}
	\label{figure:localization-simulation}
\end{figure}

The method was tested in simulation first and an example resultant localization map is shown in \cref{figure:localization-simulation}, where one jammer is simulated and surveys are performed from four locations. The likelihood of the position estimation is maximum around the position of the jammer. Notably, no threshold function is applied in this case, causing an elevated noise floor over the entire simulation area. 

\section{Experimental Setup}
\label{section:experimental_setup}

\begin{figure}[h]
	\centering
	\includegraphics[width=\columnwidth]{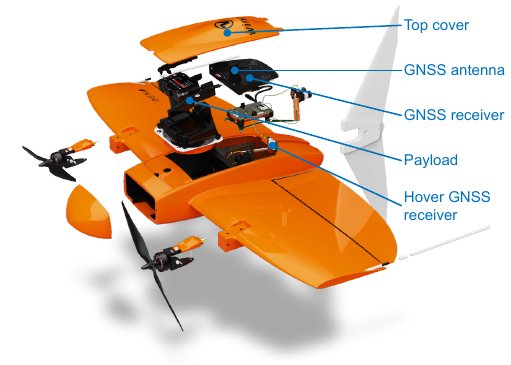}
	\caption{WingtraOne GenII VTOL survey UAV exploded view \cite{wingtraone}. }
	\label{figure:wingtra-one}
\end{figure}

The VTOL UAV used in this work is the WingtraOne GenII (\cref{figure:wingtra-one}), conventionally employed for precision land surveys and large-scale photogrammetry. The UAV has full VTOL capability and it is remotely controlled by an operator on the ground or it can fly autonomous survey missions. The payload bay hosts a survey-grade GNSS receiver and a high-resolution camera used in land surveys. For the purpose of this work, the camera is removed from the payload bay, to lighten the overall structure and improve flight time. The navigation and survey receiver are mounted together with the onboard computer inside the fuselage. The main antenna is mounted on the top of the UAV. Navigation in hover mode is provided by a forward-mounted GNSS receiver. The GNSS receiver used for navigation during conventional flight and for interference survey purposes is the Septentrio Mosaic X5 multi-band precision GNSS receiver \cite{mosaicx5}. During hovering, an additional U-Blox M9N is used to maintain position, along with the UAV's proprietary dead reckoning engine. The antenna used to perform the RF scanning is a Harxon HX-CH7018A \cite{harxon}. The GNSS navigation and survey antenna is located on the top of the fuselage and is mounted at a \SI{7}{\degree} negative angle corresponding to the angle of attack (i.e., the angle between the incoming air and the reference line of the wing) of the UAV in cruise mode. This allows for tilt compensation in navigation mode and better sensitivity towards ground-located interference sources when in survey mode. 

\begin{figure}[h]
	\centering
	\includegraphics[width=\columnwidth]{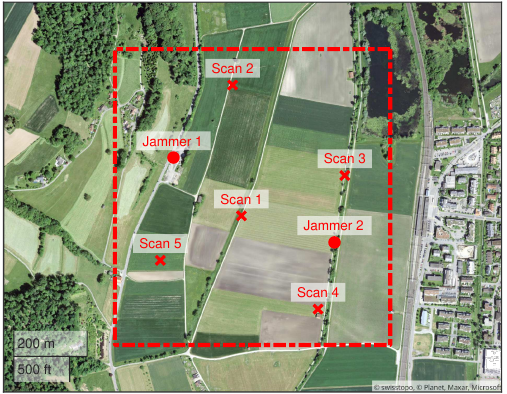}
	\caption{Test Area 1: rural environment.}
	\label{figure:test-areas-1}
\end{figure}

\begin{figure}[h]
	\centering
	\includegraphics[width=\columnwidth]{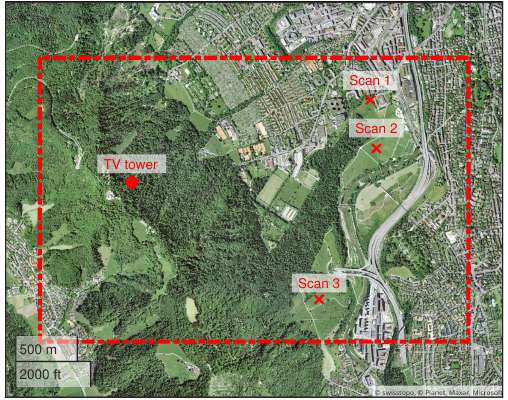}
	\caption{Test Area 2: city environment.}
	\label{figure:test-areas-2}
\end{figure}

To perform the tests, we identify two areas of $\SI{1}{\kilo\meter\squared}$ and $\SI{6}{\kilo\meter\squared}$. The selected areas are shown in \cref{figure:test-areas-1,figure:test-areas-2}. Test Area 1 (\cref{figure:test-areas-1}) is a rural environment, while Test Area 2 (\cref{figure:test-areas-2}) is a city environment rich in potential RFI transmitters such as long-range broadcast towers. The mission in each location is performed according to the "Scan 1", "Scan 2," etc., pattern shown, with a different number of scanning sites. Specifically, the mission conducted at Test Area 2 is performed near a television broadcast tower, whose location is known. The two test areas are chosen to highlight the different capabilities of the presented method. Test Area 1 (\cref{figure:test-areas-1}) allows to freely fly around and near the interference transmitters. Test Area 2 (\cref{figure:test-areas-2}) allows to test the detection capability at long range and without the possibility of obtaining vantage points all around the potential transmitter: this site shows how even in scenarios of constrained maneuverability the system is still effective.

Due to regulatory and ethical constraints, it is not possible to deploy a real GNSS jammer but we show the feasibility and capabilities of the presented method by looking at interference produced by harmonic byproducts of commercially available radio transmitters (legal to operate in the country the tests are performed) and other interference sources available in the environment.

The flight data and the output of the sensing GNSS receiver are recorded on the UAV control computer and analyzed post-flight. The post-processing of the data is divided into four parts: extraction of the navigation route and mission execution from the onboard computer; extraction of the post-processing kinematics (PPK); interference map extraction at each survey location, and fusion of survey and navigation data. Although data could be analyzed in real-time onboard the UAV, due to constraints in the available computational power, this is left for future work.

\section{Results}
\label{section:results}
\begin{figure}[h]
	\centering
	\includegraphics[width=\columnwidth]{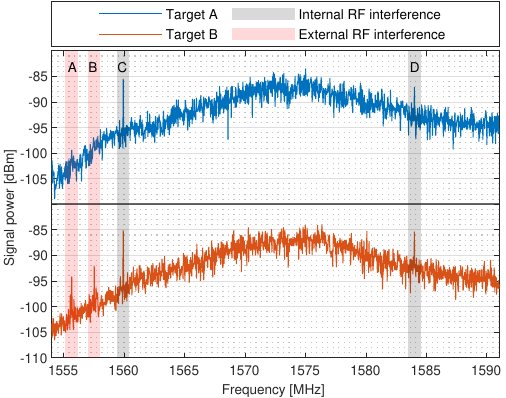}
	\caption{PSD diagram showing a capture obtained in two specific targets.}
	\label{figure:psd-waterfall}
\end{figure}

Following \cref{formula:psd}, PSDs at each location are collected during testing, of which an example from Test Area 1 is shown in \cref{figure:psd-waterfall}, in two selected heading angles. Four interference sources are visible, indicated by markers. A separate analysis conducted internally at Wingtra shows that interference occurring at markers C and D in \cref{figure:psd-waterfall} is caused by internal components of the UAV and is not relevant to the work presented here. 

\begin{figure}[h]
	\centering
	\includegraphics[width=\columnwidth]{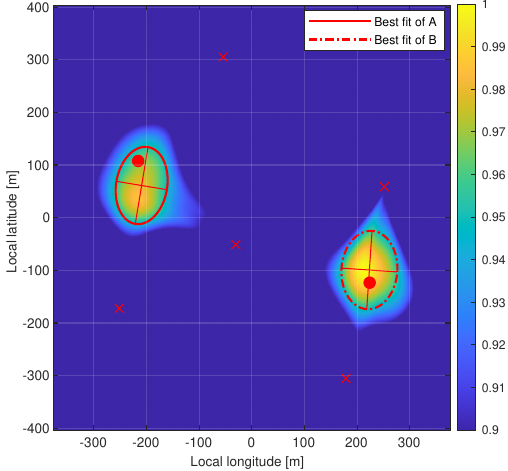}
	\caption{Test area A: two independent interference transmitters.}
	\label{figure:jammer-country}
\end{figure}

On the other hand, peaks at markers A and B are caused by external transmitters. The survey mission executed in Test Area 1 highlighted the presence of two interference transmitters located in the zones marked as Jammer 1 and Jammer 2 in \cref{figure:test-areas-1}. Further analysis of the PSD measurements shows that the center frequency of such transmitters is within harmonics multiples of the frequency allocation band for radio amateur operations. This shows the power of the presented method: although the transmitters are not directly jamming the specific GNSS frequency channel, the highlighted areas could still be affected by reduced QoS for GNSS services. Best fit ellipses of the two detected interference transmitters are provided in \cref{table:bestfit-ellipses-test-1} and \cref{figure:jammer-country}, showing both the extent of the affected area and the orientation of the transmitter antenna (in case of a directional transmitter). 

\begin{figure}[h]
	\centering
	\includegraphics[width=\columnwidth]{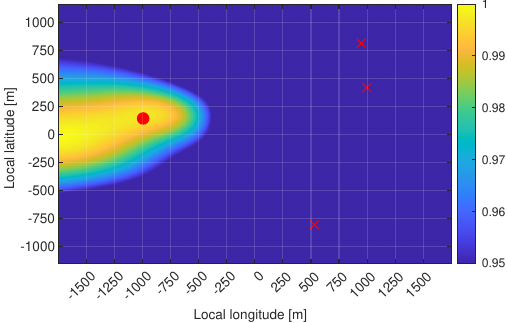}
	\caption{Test area B: spurious interference source likely due to TV tower transmissions.}
	\label{figure:jammer-tvtower}
\end{figure}

Similarly, another source of interference was detected in Test Area 2, where a TV broadcast tower was identified as a likely source of in-band spurious emissions as shown in \cref{figure:test-areas-2}. Leaky TV amplifiers are known to cause disturbances in adjacent bands due to possibly poor harmonics rejection. The scan performed from three survey locations highlights the position of the interference. It is important to notice how in this case the mission geometry is relevant. Compared to Test Area 1, the UAV mission was flown with a less favorable distribution of survey points, leading to a degenerate error ellipsis around the transmitter as shown in and \cref{figure:jammer-tvtower}. This is due to all the survey locations being on the same side of the transmitter. Best fit results are unbounded due to the poor geometry, although an elliptical best fit still provides information regarding the extent of the affected area and possibly the orientation of the transmitter (\cref{table:bestfit-ellipses-test-2}).

\begin{figure}[h]
	\centering
	\includegraphics[width=0.9\columnwidth]{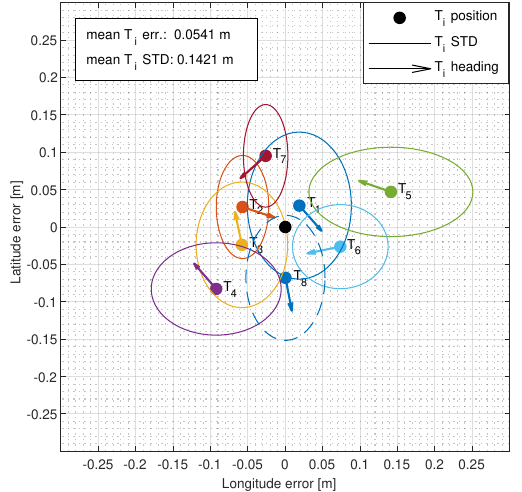}
	\caption{Position holding accuracy during survey measurements (INS+GNSS aiding).}
	\label{figure:position-holding}
\end{figure}

\begin{figure}[h]
	\centering
	\includegraphics[width=\columnwidth]{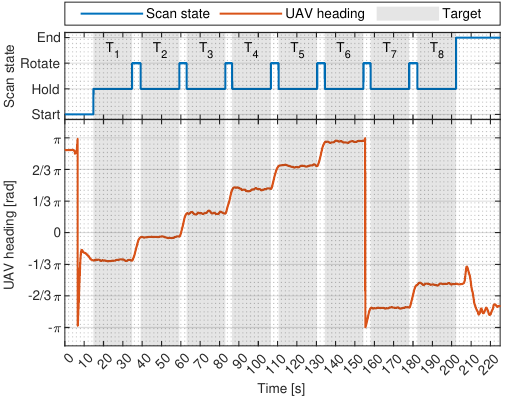}
	\caption{Antenna heading change at a given scan position (absolute heading to true north).}
	\label{figure:beam-steering}
\end{figure}

Several missions flown in different weather conditions show that results are critically dependent on the crosswind at the location of the survey. In good weather conditions, with crosswinds below \SI{3}{\meter/\second}, the INS-aided GNSS position holding shows an average \SI{0.05}{\meter} position holding error with a standard deviation of \SI{0.142}{\meter}, with accurate antenna heading positioning as shown in \cref{figure:position-holding}.  For each transition time ($T_i$) corresponding to a specific heading angle, as shown in \cref{figure:beam-steering}, the position accuracy and heading of the survey antenna are shown in \cref{figure:position-holding}. The position used to reference the individual spectrum measurement is obtained as an average of the positions at each heading and.
Progressive degradation was observed for higher wind speeds, especially when the wind direction is normal to the UAV's plane, which is the major limitation of the VTOL system. This is shown in the error ellipsis in \cref{figure:position-holding}, which grows larger in the direction the wind is pushing the UAV. 

\begin{table}[h]
\renewcommand{\arraystretch}{1.1}
\centering
\begin{tabularx}{\columnwidth} { 
  | >{\raggedright\arraybackslash}X 
  | >{\raggedleft\arraybackslash}X 
  | >{\raggedleft\arraybackslash}X | }\hline
\bfseries  & \bfseries Best Fit A & \bfseries Best Fit B \\
\hline\hline
Long Axis                & \SI{147.94}{\meter} & \SI{149.00}{\meter}\\\hline
Short Axis               & \SI{97.05}{\meter} & \SI{106.05}{\meter}\\\hline
Heading on Local northing & \SI{9.40}{\degree} & \SI{3.86}{\degree}\\\hline
Local Easting            & \SI{-209.30}{\meter} & \SI{223.21}{\meter}\\\hline
Local Northing           & \SI{61.08}{\meter} & \SI{-99.41}{\meter}\\   
\hline
\end{tabularx}
\caption{\textbf{Summary of the best fit:} localization and extent of the affected zones in Test Area 1.}
\label{table:bestfit-ellipses-test-1}
\end{table}

\begin{table}[h]
\renewcommand{\arraystretch}{1.1}
\centering
\begin{tabularx}{\columnwidth} { 
  | >{\raggedright\arraybackslash}X 
  | >{\raggedleft\arraybackslash}X 
  | >{\raggedleft\arraybackslash}X | }\hline
\bfseries  & \bfseries Best Fit A  \\
\hline\hline
Long Axis                & unbound \\\hline
Short Axis               & \SI{480.21}{\meter} \\\hline
Heading on Local northing & \SI{-12.78}{\degree} \\\hline
Local Easting (Focus)           & \SI{-1150.70}{\meter} \\\hline
Local Northing (Focus)          & \SI{114.78}{\meter} \\   
\hline
\end{tabularx}
\caption{\textbf{Summary of the best fit:} localization and extent of the affected zones in Test Area 2.}
\label{table:bestfit-ellipses-test-2}
\end{table}

\textbf{Measurement uncertainties and operation in a GNSS denied area} - The errors analyzed in \cref{table:bestfit-ellipses-test-1,table:bestfit-ellipses-test-2} are not to be traced to the position holding in hover mode, which shows decimeter-level accuracy. On the other hand, the most likely source of uncertainty in the system comes from the shape of the antenna pattern. Given that the antenna is used in cruise mode to provide precise positioning, it is not advisable to use an antenna with a narrower radiation pattern which would be beneficial for the interference mapping task, given that this would lead to a reduced localization accuracy. On the other hand, a wider radiation pattern allows for faster measuring of the interference in hover mode, extending the duration of the mission. Further tests and experiments to optimize the trade-off are ongoing.

The algorithm simultaneously maps the extent of the affected area and estimates the location of the adversarial transmitter. This allows for early detection and estimation of the extent of the denied area. With this knowledge, the UAV can avoid the interference area: by navigating at the edge of the interference-affected area, future work will focus on updating the mission path to simultaneously complete the designed mission and avoid the interference areas, as a preemptive awareness tool. While it is possible, for short periods, for the UAV to rely on the inertial navigation system to reliably fly, this should only be used to quickly evade a GNSS-denied area. In particular, the case in which the start of operation is within the jammed area is beyond the current scope of the work and can be solved by applying pre-flight detection methods which are beyond the scope of this work.

Notably, the system measures the amount of in-band interference for each relevant GNSS frequency channel, without considering the interference's actual effect on the GNSS signal. Nevertheless, it is important to stress that the method presented here is to be used preemptively: for safety and robustness reasons, the VTOL should avoid operating within jammed areas. The high sensitivity of the antenna combined with the methods presented in this work should provide sufficient awareness of the RFI situation so that the mission path can be modified accordingly to avoid problematic areas.

While the presented method is capable of detecting multiple interference sources, transmitters that are close to each other might lead to a degenerate localization where two RFI sources are marked as one. Similarly, in case the interference source is located close to the survey points the resolution is limited by the "field of view" of the antenna, with power detected at multiple headings. While the former might not be an issue, as it would simply lead to a larger area of possible interference, the second issue directly limits the localization accuracy.  

The detection of a multiplicity of possibly colluding sources represents a difficult problem, more so in a complex multipath environment, where reflections might trigger the detection of mirror transmitters (e.g., the localization of the reflected source rather than the actual transmitter). While further investigation is necessary, a possible solution is to increase the number of survey locations to avoid echoes. Notably, all scans are performed by disabling the receiver's Automatic Gain Control (AGC). This approach proved to be beneficial to avoid gain compression due to the receiver avoiding saturation of the front-end. While this is important to ensure optimal sensitivity, it is also undesirable for sensing purposes. While the presented work does not rely on absolute power measurements, the relative signal strength of the interference sources is important to extract the expectation density map.
	
\section{Conclusions and future work}
\label{section:conclusions}

This work presented a novel approach to long-range RFI survey missions, leveraging optimized flight dynamics using VTOL planes. The presented interference localization method based on scans from distributed vantage points proved capable of isolating multiple RFI transmitters on the ground and spurious emitters located above the ground (given the transmitter of the TV tower antenna is located about \SI{20}{\meter} above the base of the antenna mast), which are possibly even more effective in disrupting GNSS signals reception. Leveraging the internal GNSS receiver used for navigation in cruise flight mode avoids deploying additional hardware, relying on the available GNSS receiver front-end for relative power measurements. On the other hand, operations in hovering mode are highly power-consuming, somewhat limiting the number of scans that can be performed during a single mission. Nevertheless, the optimal cruise mode allows for a larger operational area and significantly better geometrical dispersion of the survey points, allowing the flight planner to select fewer, more convenient points for scanning.

Results under real-world jamming conditions and devices and will be presented in a follow-up, based on data collected at Jammertest 2023 (which includes actual jammers) to evaluate the robustness of the presented platform to detect and isolate commonly used PPD equipment. Additionally, extensions toward classification of the interference source and estimation of the impact on the GNSS receiver QoS are in development.  Remote sensing capabilities are important to modify in real-time the flight mission to avoid GNSS-denied areas overall improving the robustness and safety of the system, which is specifically relevant for unmanned aerial platforms.  Combined with on-board processing of the measurements, this approach would provide full situational awareness to the control system to avoid GNSS-denied areas while being able to complete the required task.

\acknowledgments
This work was supported in part by the SSF SURPRISE project and the Security Link strategic research center.
Additionally, the authors thank Wim De Wilde from Septentrio N.V. for the invaluable discussions. His contribution of ideas and suggestions provided crucial perspectives to our work.

\bibliographystyle{IEEEtran}
\bibliography{bibfile}

\begin{thebibliography}{10}
\providecommand{\url}[1]{#1}
\csname url@samestyle\endcsname
\providecommand{\newblock}{\relax}
\providecommand{\bibinfo}[2]{#2}
\providecommand{\BIBentrySTDinterwordspacing}{\spaceskip=0pt\relax}
\providecommand{\BIBentryALTinterwordstretchfactor}{4}
\providecommand{\BIBentryALTinterwordspacing}{\spaceskip=\fontdimen2\font plus
\BIBentryALTinterwordstretchfactor\fontdimen3\font minus
  \fontdimen4\font\relax}
\providecommand{\BIBforeignlanguage}[2]{{%
\expandafter\ifx\csname l@#1\endcsname\relax
\typeout{** WARNING: IEEEtran.bst: No hyphenation pattern has been}%
\typeout{** loaded for the language `#1'. Using the pattern for}%
\typeout{** the default language instead.}%
\else
\language=\csname l@#1\endcsname
\fi
#2}}
\providecommand{\BIBdecl}{\relax}
\BIBdecl

\bibitem{psiaki2016gnss}
M.~L. Psiaki and T.~E. Humphreys, ``{GNSS} spoofing and detection,''
  \emph{Proceedings of the IEEE}, vol. 104, no.~6, pp. 1258--1270, 2016.

\bibitem{Lenhart2022}
M.~Lenhart, M.~Spanghero, and P.~Papadimitratos, ``{Distributed and Mobile
  Message Level Relaying/Replaying of GNSS Signals},'' in \emph{2022
  International Technical Meeting of The Institute of Navigation (ITM)}, Long
  Beach, California, 2022, pp. 56--57.

\bibitem{HumphreysAssessingSpoofer}
T.~E. Humphreys, B.~M. Ledvina, M.~L. Psiaki \emph{et~al.}, ``Assessing the
  spoofing threat: Development of a portable gps civilian spoofer,'' in
  \emph{21st International Technical Meeting of the Satellite Division of The
  Institute of Navigation (ION GNSS 2008)}, Savannah, Georgia, 1987, pp.
  2314--2325.

\bibitem{CNETNewark}
\BIBentryALTinterwordspacing
``{Truck driver has GPS jammer, accidentally jams Newark airport - CNET}.''
  [Online]. Available:
  \url{https://www.cnet.com/culture/truck-driver-has-gps-jammer-accidentally-jams-newark-airport}
\BIBentrySTDinterwordspacing

\bibitem{insideGNSSNewark}
\BIBentryALTinterwordspacing
``{FCC Fines Operator of GPS Jammer That Affected Newark Airport GBAS - Inside
  GNSS}.'' [Online]. Available:
  \url{https://insidegnss.com/fcc-fines-operator-of-gps-jammer-that-affected-newark-airport-gbas}
\BIBentrySTDinterwordspacing

\bibitem{SkytruthJamming}
\BIBentryALTinterwordspacing
``{Systematic GPS Manipulation Occuring at Chinese Oil Terminals and Government
  Installations – SkyTruth}.'' [Online]. Available:
  \url{https://skytruth.org/2019/12/systematic-gps-manipulation-occuring-at-chinese-oil-terminals-and-government-installations}
\BIBentrySTDinterwordspacing

\bibitem{john2001vulnerability}
J.~Volpe, ``{Vulnerability assessment of the transportation infrastructure
  relying on the global positioning system},'' \emph{National Transportation
  Systems Center}, 2001.

\bibitem{Mitch2011}
R.~H. Mitch, R.~C. Dougherty, M.~L. Psiaki, S.~P. Powell, B.~W. O'Hanlon, J.~A.
  Bhatti, and T.~E. Humphreys, ``{Signal characteristics of civil GPS
  jammers},'' in \emph{24th International Technical Meeting of the Satellite
  Division of the Institute of Navigation 2011, ION GNSS 2011}, vol.~3, 2011,
  pp. 1907--1919.

\bibitem{Borio2012}
D.~Borio, C.~O'Driscoll, and J.~Fortuny, ``{GNSS jammers: Effects and
  countermeasures},'' in \emph{6th ESA Workshop on Satellite Navigation
  Technologies: Multi-GNSS Navigation Technologies Galileo's Here, NAVITEC 2012
  and European Workshop on GNSS Signals and Signal Processing}, 2012.

\bibitem{Anritsu_appnote_rfi}
\BIBentryALTinterwordspacing
Anritsu, ``{Spotting Interference or What Am I looking for? Spotting
  Interference in the Field}.'' [Online]. Available:
  \url{https://dl.cdn-anritsu.com/en-us/test-measurement/files/Application-Notes/Application-Note/11410-00972A.pdf}
\BIBentrySTDinterwordspacing

\bibitem{Keysight_appnote_rfi}
\BIBentryALTinterwordspacing
Keysight, ``{Techniques for Precise Measurement Calibrations in the Field}.''
  [Online]. Available:
  \url{https://www.keysight.com/us/en/assets/7018-03476/application-notes/5991-0418.pdf}
\BIBentrySTDinterwordspacing

\bibitem{Isoz2010}
O.~Isoz, A.~T. Balaei, and D.~M. Akos, ``{Interference detection and
  localization in GPS L1 band},'' in \emph{Institute of Navigation -
  International Technical Meeting 2010, ITM 2010}, vol.~2, jan 2010, pp.
  1095--1099.

\bibitem{olsson2022participatory}
G.~K. Olsson, E.~Axell, E.~G. Larsson, and P.~Papadimitratos, ``Participatory
  sensing for localization of a gnss jammer,'' in \emph{2022 International
  Conference on Localization and GNSS (ICL-GNSS)}, 2022, pp. 1--7.

\bibitem{olsson2023participatory}
G.~K. Olsson, S.~Nilsson, E.~Axell, E.~G. Larsson, and P.~Papadimitratos,
  ``Using mobile phones for participatory detection and localization of a gnss
  jammer,'' in \emph{2023 IEEE/ION Position, Location and Navigation Symposium
  (PLANS)}, 2023, pp. 536--541.

\bibitem{Lindstrom2007}
J.~Lindstr{\"{o}}m, D.~M. Akos, O.~Isoz, and M.~Junered, ``{GNSS interference
  detection and localization using a network of low cost front-end modules},''
  in \emph{20th International Technical Meeting of the Satellite Division of
  The Institute of Navigation 2007 ION GNSS 2007}, vol.~4, sep 2007, pp.
  1165--1172.

\bibitem{Strizic2018}
L.~Strizic, D.~Akos, and S.~Lo, ``{Crowdsourcing GNSS JammingDetection and
  Localization},'' in \emph{Proceedings of the 2018 International Technical
  Meeting of The Institute of Navigation, ITM 2018}, vol. 2018-Janua.\hskip 1em
  plus 0.5em minus 0.4em\relax Institute of Navigation, feb 2018, pp. 1--33.

\bibitem{Cetin2014}
E.~Cetin, R.~J. Thompson, M.~Trinkle, and A.~G. Dempster, ``{Interference
  detection and localization within the GNSS Environmental Monitoring System
  (GEMS) - System update and latest field test results},'' in \emph{27th
  International Technical Meeting of the Satellite Division of the Institute of
  Navigation, ION GNSS 2014}, vol.~4, 2014, pp. 3449--3460.

\bibitem{GisdakisGP:J:2016}
S.~Gisdakis, T.~Giannetsos, and P.~Papadimitratos, ``{Security, Privacy, and
  Incentive Provision for Mobile Crowd Sensing Systems},'' \emph{IEEE Internet
  of Things Journal}, vol.~3, no.~5, pp. 839--853, October 2016.

\bibitem{GisdakisGP:C:2015}
------, ``{SHIELD: A Data Verification Framework for Participatory Sensing
  Systems},'' in \emph{{ACM} Conference on Security {\&} Privacy in Wireless
  and Mobile Networks {(ACM WiSec)}}, New York, NY, USA, June 2015, pp.
  16:1--16:12.

\bibitem{Poncelet2012}
J.~P. Poncelet and D.~M. Akos, ``{A low-cost monitoring station for detection
  \& localization of interference in GPS L1 band},'' in \emph{6th ESA Workshop
  on Satellite Navigation Technologies: Multi-GNSS Navigation Technologies
  Galileo's Here, NAVITEC 2012 and European Workshop on GNSS Signals and Signal
  Processing}, 2012.

\bibitem{Akos2012}
D.~M. Akos, ``{Who's afraid of the spoofer? GPS/GNSS spoofing detection via
  automatic gain control (agc)},'' \emph{Navigation, Journal of the Institute
  of Navigation}, vol.~59, no.~4, pp. 281--290, dec 2012.

\bibitem{Spens2022}
N.~Spens, D.~K. Lee, F.~Nedelkov, and D.~Akos, ``{Detecting GNSS Jamming and
  Spoofing on Android Devices},'' \emph{Navigation, Journal of the Institute of
  Navigation}, vol.~69, no.~3, sep 2022.

\bibitem{Hashemi2019}
``{STRIKE3-Case Study for Standardized Testing of Timing-Grade GNSS Receivers
  Against Real-World Interference Threats},'' in \emph{2019 International
  Conference on Localization and GNSS (ICL-GNSS)}.\hskip 1em plus 0.5em minus
  0.4em\relax IEEE, jun 2019, pp. 1--8.

\bibitem{Montgomery2009}
P.~Y. Montgomery, T.~E. Humphreys, and B.~M. Ledvina, ``{Receiver-autonomous
  spoofing detection: Experimental results of a multi-antenna receiver defense
  against a portable civil GPS spoofer},'' in \emph{Proceedings of the
  Institute of Navigation, National Technical Meeting}, vol.~1, 2009, pp.
  124--130.

\bibitem{Magiera2015}
J.~Magiera and R.~Katulski, ``{Detection and mitigation of GPS spoofing based
  on antenna array processing},'' \emph{Journal of Applied Research and
  Technology}, vol.~13, no.~1, pp. 45--57, feb 2015.

\bibitem{Bhatti2012}
J.~A. Bhatti, T.~E. Humphreys, and B.~M. Ledvina, ``{Development and
  demonstration of a TDOA-based GNSS interference signal localization
  system},'' in \emph{Record - IEEE PLANS, Position Location and Navigation
  Symposium}, 2012, pp. 455--469.

\bibitem{Bours2014}
A.~Bours, E.~Cetin, and A.~G. Dempster, ``{Enhanced GPS interference detection
  and localisation},'' \emph{Electronics Letters}, vol.~50, no.~19, pp.
  1391--1393, sep 2014.

\bibitem{Caizzone2019}
S.~Caizzone, G.~Buchner, M.~S. Circiu, M.~Cuntz, W.~Elmarissi, and E.~P.
  Marcos, ``{A miniaturized multiband antenna array for robust navigation in
  aerial applications},'' \emph{Sensors (Switzerland)}, vol.~19, no.~10, may
  2019.

\bibitem{Perez-Marcos2023}
E.~Perez-Marcos, M.~Cuntz, A.~Konovaltsev, L.~Kurz, S.~Caizzone, and M.~Meurer,
  ``{CRPA and Array Receivers for Civil GNSS Applications},'' in \emph{2023
  IEEE/ION Position, Location and Navigation Symposium, PLANS 2023}.\hskip 1em
  plus 0.5em minus 0.4em\relax Institute of Electrical and Electronics
  Engineers Inc., 2023, pp. 318--328.

\bibitem{Marcos2018}
E.~P. Marcos, A.~Konovaltsev, S.~Caizzone, M.~Cuntz, K.~Yinusa, W.~Elmarissi,
  and M.~Meurer, ``{Interference and spoofing detection for GNSS maritime
  applications: Using direction of arrival and conformal antenna array},'' in
  \emph{Proceedings of the 31st International Technical Meeting of the
  Satellite Division of the Institute of Navigation, ION GNSS+ 2018}.\hskip 1em
  plus 0.5em minus 0.4em\relax Institute of Navigation, 2018, pp. 2907--2922.

\bibitem{Clements2022}
Z.~Clements, P.~Ellis, M.~Psiaki, and T.~E. Humphreys, ``{Geolocation of
  Terrestrial GNSS Spoofing Signals from Low Earth Orbit},'' in
  \emph{Proceedings of the 35th International Technical Meeting of the
  Satellite Division of The Institute of Navigation (ION GNSS+ 2022)}.\hskip
  1em plus 0.5em minus 0.4em\relax Institute of Navigation, sep 2022, pp.
  3418--3431.

\bibitem{Clements2023}
Z.~Clements, T.~E. Humphreys, and P.~Ellis, ``{Dual-Satellite Geolocation of
  Terrestrial GNSS Jammers from Low Earth Orbit},'' in \emph{2023 IEEE/ION
  Position, Location and Navigation Symposium, PLANS 2023}.\hskip 1em plus
  0.5em minus 0.4em\relax Institute of Electrical and Electronics Engineers
  Inc., 2023, pp. 458--469.

\bibitem{Spicer2015}
J.~Spicer, A.~Perkins, L.~Dressel, M.~James, Y.~H. Chen, D.~S. {De Lorenzo},
  and P.~Enge, ``{The JAGER project: GPS Jammer hunting with a multi-purpose
  UAV test platform},'' \emph{Institute of Navigation International Technical
  Meeting 2015, ITM 2015}, pp. 62--70, 2015.

\bibitem{Perkins2016}
A.~Perkins, L.~Dressel, S.~Lo, T.~Reid, K.~Gunning, and P.~Enge,
  ``{Demonstration of UAV-Based GPS jammer localization during a live
  interference exercise},'' \emph{29th International Technical Meeting of the
  Satellite Division of the Institute of Navigation, ION GNSS 2016}, vol.~5,
  pp. 3094--3106, 2016.

\bibitem{Perkins2019}
A.~Perkins, Y.~H. Chen, S.~Lo, C.~Lee, and J.~{David Powell}, ``{Real-time
  unmanned aerial system (UAS) based interference localization in a GNSS denied
  environment},'' in \emph{Proceedings of the 32nd International Technical
  Meeting of the Satellite Division of the Institute of Navigation, ION GNSS+
  2019}.\hskip 1em plus 0.5em minus 0.4em\relax Institute of Navigation, sep
  2019, pp. 1003--1019.

\bibitem{matrice600}
\BIBentryALTinterwordspacing
DJI, ``{DJI MATRICE 600}.'' [Online]. Available:
  \url{https://www.dji.com/se/matrice600}
\BIBentrySTDinterwordspacing

\bibitem{PTSSpectrum}
\BIBentryALTinterwordspacing
``{Spectrum Strategy and Spectrum Orientation Plan - Swedish Post and Telecom
  Authority}.'' [Online]. Available:
  \url{https://www.pts.se/en/english-b/radio/spectrum-policy-and-spectrum-orientation-plan}
\BIBentrySTDinterwordspacing

\bibitem{Curran2016}
\BIBentryALTinterwordspacing
J.~T. Curran, M.~Bavaro, P.~Closas, and M.~Navarro, ``On the threat of
  systematic jamming of gnss,'' in \emph{ION GNSS+, The International Technical
  Meeting of the Satellite Division of The Institute of Navigation}, ser. GNSS
  2016.\hskip 1em plus 0.5em minus 0.4em\relax Institute of Navigation, Nov.
  2016. [Online]. Available: \url{http://dx.doi.org/10.33012/2016.14672}
\BIBentrySTDinterwordspacing

\bibitem{Miguel2023}
N.~R.~S. Miguel, Y.~H. Chen, S.~Lo, T.~Walter, and D.~Akos, ``{Calibration of
  RFI Detection Levels in a Low-Cost GNSS Monitor},'' in \emph{2023 IEEE/ION
  Position, Location and Navigation Symposium, PLANS 2023}.\hskip 1em plus
  0.5em minus 0.4em\relax Institute of Electrical and Electronics Engineers
  Inc., 2023, pp. 520--535.

\bibitem{mosaicx5}
\BIBentryALTinterwordspacing
``{Septentrio Mosaic X5 Product Datasheet - Septentrio}.'' [Online]. Available:
  \url{https://www.septentrio.com/en/products/gps/gnss-receiver-modules/mosaic-x5}
\BIBentrySTDinterwordspacing

\bibitem{harxon}
\BIBentryALTinterwordspacing
``D-helix antenna hx-ch7018a - harxon.'' [Online]. Available:
  \url{https://en.harxon.com/product/detail/133}
\BIBentrySTDinterwordspacing

\bibitem{wingtraone}
\BIBentryALTinterwordspacing
``{WingtraOne GenII Technical Specification - Wingtra}.'' [Online]. Available:
  \url{https://wingtra.com/wp-content/uploads/Wingtra-Technical-Specifications.pdf}
\BIBentrySTDinterwordspacing

\end{thebibliography}


\newpage

\thebiography
\begin{biographywithpic}
{Marco Spanghero}{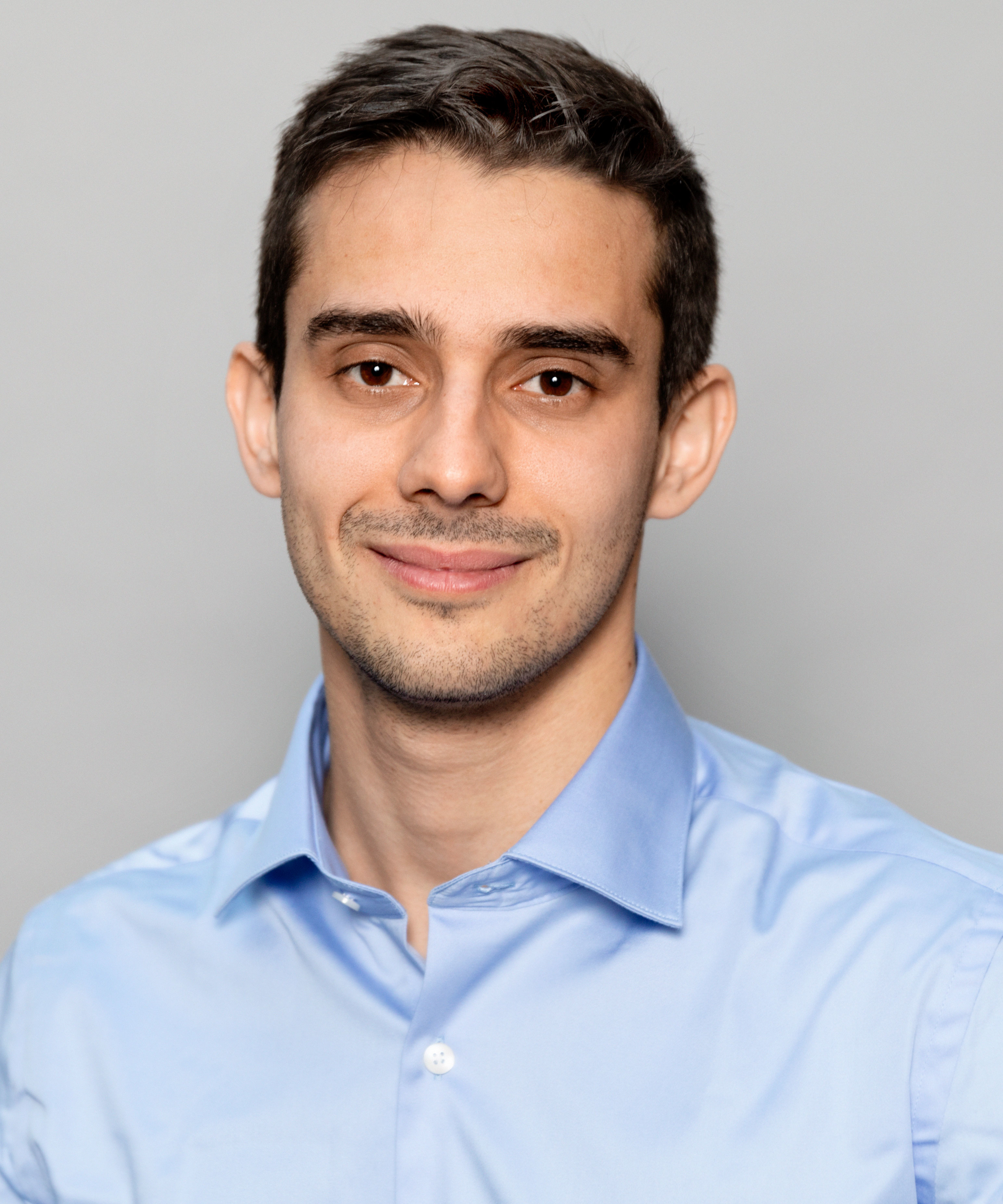}
received his B.S. from Politecnico of Milano and an MSc degree from KTH Royal Institute of Technology, Stockholm, Sweden. He is currently a Ph.D. candidate with the Networked Systems Security (NSS) group at KTH, Stockholm, Sweden, and associate with the WASP program from the Knut and Alice Wallenberg Foundation. 
\end{biographywithpic} 

\begin{biographywithpic}
{Filip Geib}{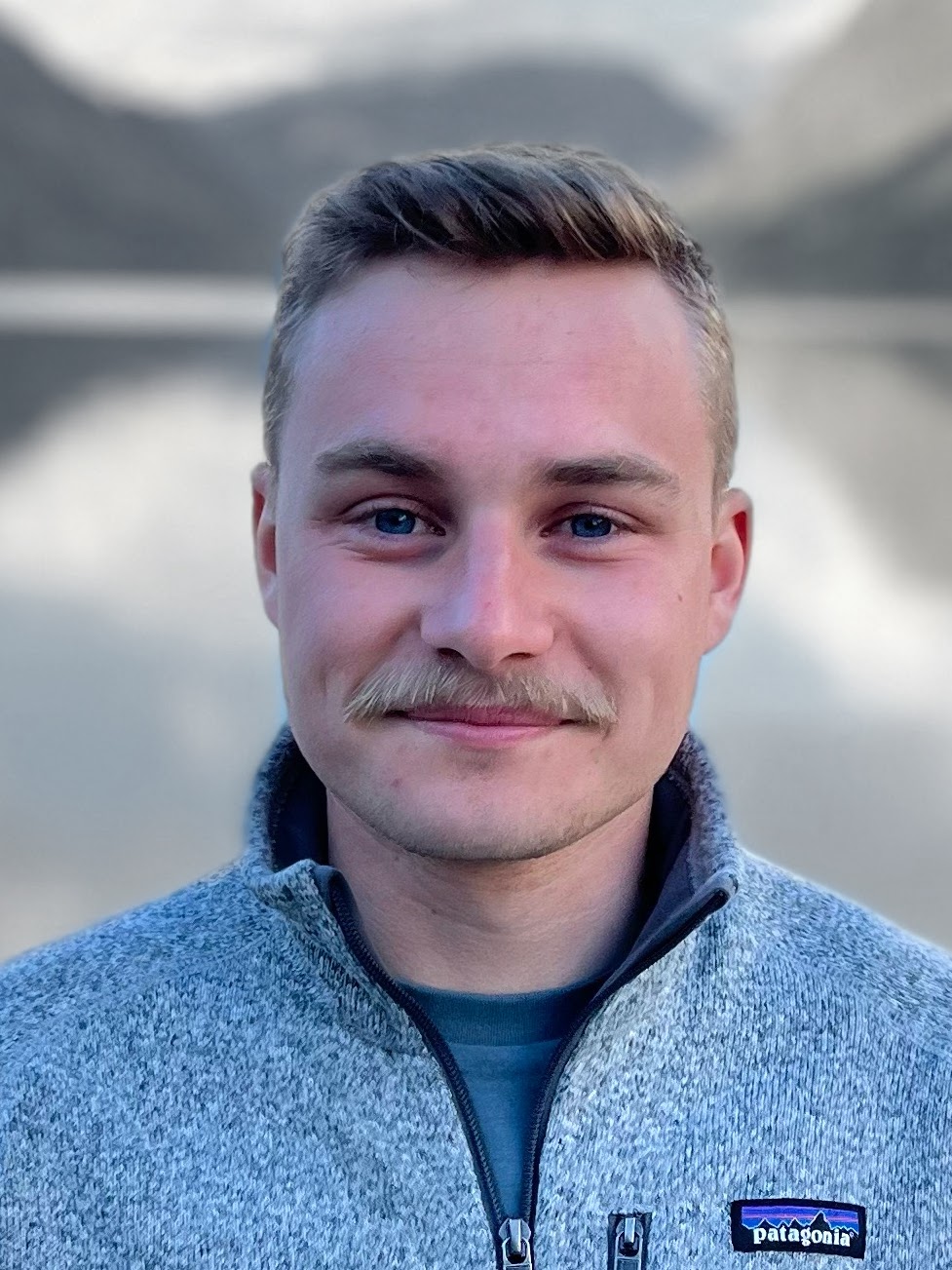}
earned his B.C. degree in Cybernetics and Robotics from the Czech Technical University in Prague in 2021.
Presently, he contributes as a Robotics Software Engineer at Wingtra AG. \\ \\
\end{biographywithpic}

\begin{biographywithpic}
{Ronny Pannier}{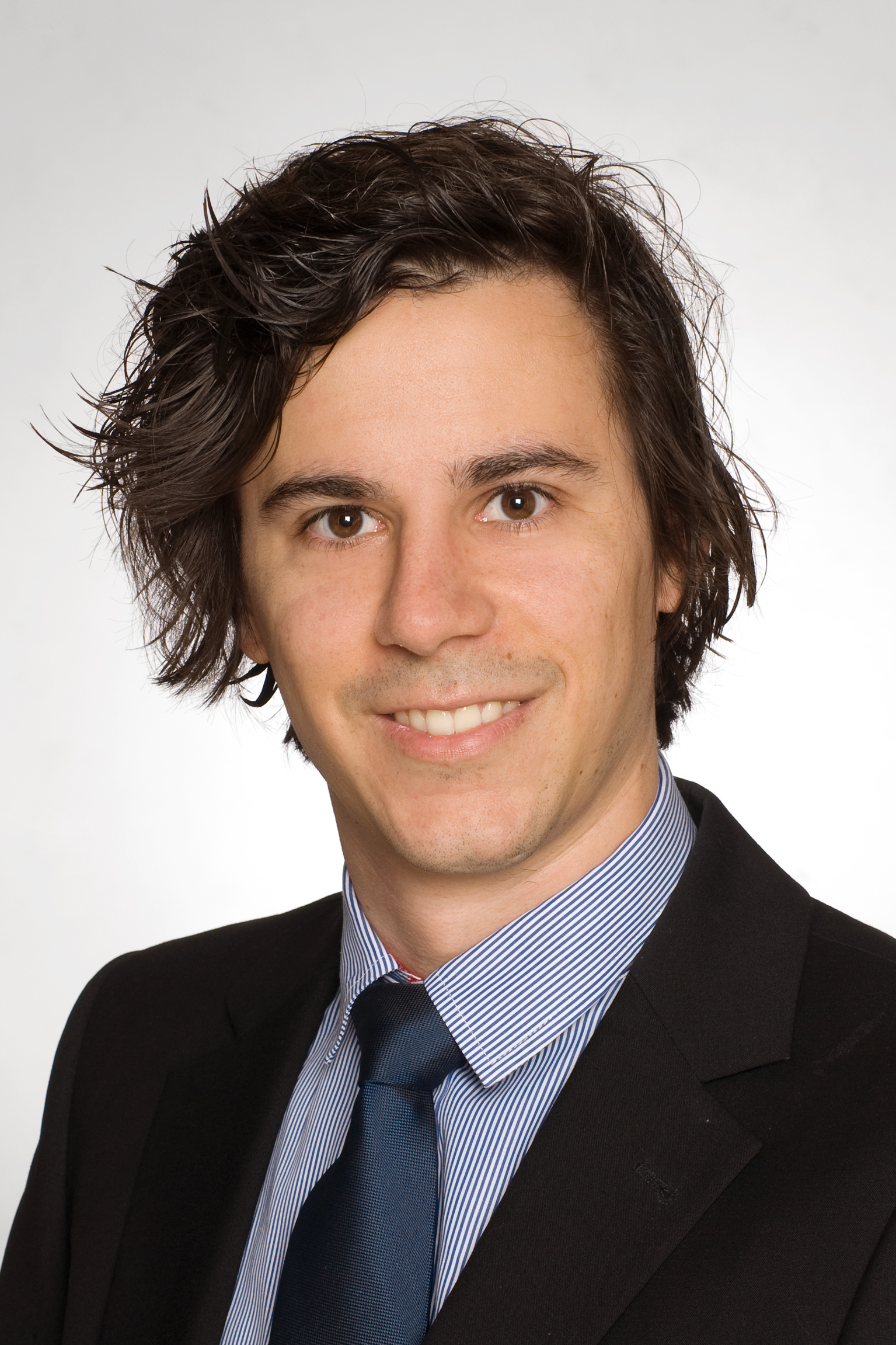}
received his MSc degree in Mechanical Engineering in 2015 from ETH Zurich in Switzerland. He's currently working as a team-lead Systems Engineering at Wingtra AG. \\ \\ \\
\end{biographywithpic}

\begin{biographywithpic}
{Panos Papadimitratos}{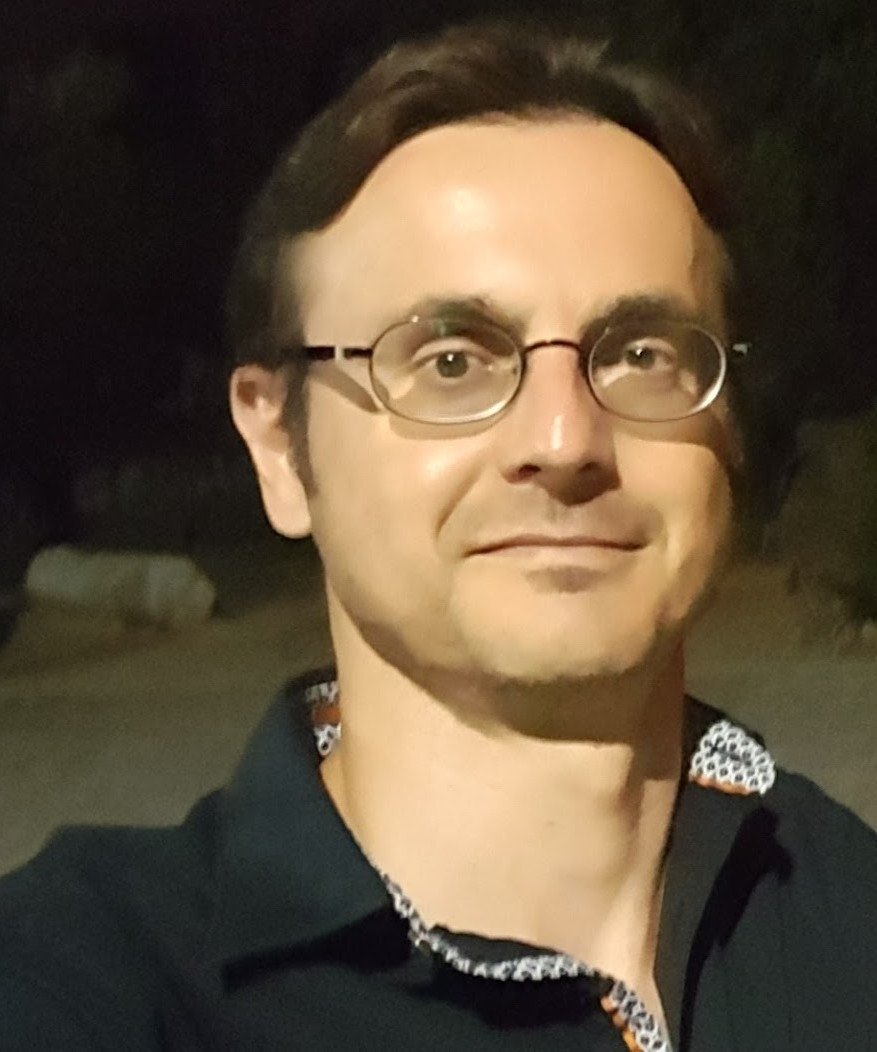}
(Fellow, IEEE) earned his Ph.D. degree from Cornell University, Ithaca, NY, USA. At KTH, Stockholm, Sweden, he leads the Networked Systems Security (NSS) group and he is a member of the Steering Committee of the Security Link Center. He serves or served as member of the ACM WiSec and CANS conference steering committees and the PETS Editorial and Advisory Boards; Program Chair for the ACM WiSec’16, TRUST’16, and CANS’18 conferences; General Chair for the ACM WISec’18, PETS’19, and IEEE EuroS\&P’19 conferences; Associate Editor of the IEEE TMC, IEEE/ACM ToN and IET IFS journals, and Chair of the Caspar Bowden PET Award. Panos is a Fellow of the Young Academy of Europe, a Knut and Alice Wallenberg Academy Fellow, and an ACM Distinguished Member. NSS webpage: \url{https://www.eecs.kth.se/nss}.
\end{biographywithpic}

\end{document}